\documentclass[12pt]{iopart}
\usepackage{epsfig,url,cite}  
\usepackage[dvips]{color}
\definecolor{Black}{named}{Black}
\definecolor{Red}{named}{Red}
\definecolor{Blue}{named}{Blue}

\def\be{\begin{equation}}
\def\ee{\end{equation}}
\def\ba{\begin{eqnarray}}
\def\ea{\end{eqnarray}}

\def\lsim{\raise0.3ex\hbox{$\;<$\kern-0.75em\raise-1.1ex\hbox{$\sim\;$}}}
\def\gsim{\raise0.3ex\hbox{$\;>$\kern-0.75em\raise-1.1ex\hbox{$\sim\;$}}}

\def\eps{\varepsilon}
\def\theta{\vartheta}

\def\vx{{\bf x}}
\def\vp{{\bf p}}

\def\d{{\rm d}}

\def\ap{\approx}

\begin{document}

\title[High energy radiation from Cen A]{High energy radiation from Centaurus A}
\author{M.~Kachelrie{\ss}$^1$, S.~Ostapchenko$^{1,2}$ and R.~Tom\`as$^3$}
\address{
$^1$Institutt for fysikk, NTNU Trondheim, Norway}
\address{$^2$D.~V.~Skobeltsyn Institute of Nuclear Physics,
 Moscow State University, Russia}
\address{$^3$II. Institut f\"ur Theoretische Physik, 
Universit\"at Hamburg, Germany}

\begin{abstract}
We calculate for the nearest active galactic nucleus (AGN), Centaurus~A, the 
flux of high energy cosmic rays and of accompanying secondary 
photons and neutrinos expected from hadronic interactions in the source. 
We use as two basic models for the generation of ultrahigh energy  cosmic 
rays (UHECR) shock acceleration in the radio jet and acceleration in the 
regular electromagnetic field close to the core of the AGN. While scattering 
on photons dominates in scenarios with acceleration close to the core, 
scattering on gas becomes more important if acceleration takes place along 
the jet. 
Normalizing the UHECR flux from Centaurus~A to the observations of the Auger 
experiment, the neutrino flux may be marginally observable in a 1\,km$^3$ 
neutrino telescope, if a steep UHECR flux $\d N/\d E\propto E^{-\alpha}$ with 
$\alpha=2.7$  extends down to $10^{17}$\,eV. 
The associated photon flux is close to or exceeds the observational data of 
atmospheric Cherenkov and $\gamma$-ray telescopes for $\alpha\gsim 2$. 
In particular, we find that already present data favour either a softer 
UHECR injection spectrum than $\alpha=2.7$ for Centaurus~A or a lower 
UHECR flux than expected from the normalization to the Auger observations.
\end{abstract}



\section{Introduction}

Progress in cosmic ray (CR) physics has been hampered for long time by
the deflection of charged cosmic rays in magnetic fields, preventing the 
identification of individual sources. Numerous searches for anisotropies 
and correlation studies have been performed, without reaching unanimous
conclusions~\cite{CR}. 
Using neutral messengers that should be produced as secondaries in 
proton-photon and proton-proton interactions close to the source  
for the identification of the sources has its own problems: 
First, secondary photons generated by hadronic CR interactions are difficult 
to disentangle from photons produced by synchrotron radiation
or inverse Compton scattering of electrons. Moreover, high energy photons 
are strongly absorbed both in the source and propagating over extragalactic 
distances. By
contrast, the extremely large mean free path of neutrinos together with the 
relatively poor angular resolution of neutrino telescopes ($\sim 1^\circ$) 
and the small expected event numbers makes the identification of 
extragalactic sources  challenging using only the neutrino 
signal. Performing neutrino astronomy beyond the establishment 
of a diffuse neutrino background requires therefore most likely additional 
input, either timing or angular information from high energy photon or CR 
experiments.

The recently announced evidence~\cite{pao_corr} for a correlation of the 
arrival directions of ultrahigh energy (UHE) CRs observed by the Pierre Auger 
Observatory (PAO) with active galactic nuclei (AGN) may provide a first test 
case for successful ``multi-messenger astronomy.'' In particular, 
Ref.~\cite{pao_corr} 
finds two events within the search bin of $3.1^\circ$ around the nearest 
active galaxy, Centaurus~A (Cen~A or NGC~5128), that is located close to 
the supergalactic plane. At present 
this correlation has only $3\,\sigma$~C.L. and other source types that 
follow the large-scale structure of matter would also result in an 
excess of events along the supergalactic plane. Independent evidence for the 
AGN source hypothesis is the characteristic bias of AGN with respect to the 
large-scale structure that is indeed reflected in the angular distribution 
of the observed UHECR 
arrival directions~\cite{ATF}. On the other hand, the correlation with
AGN is not confirmed by the data from the HiRes experiment~\cite{HiRes}. 
It is therefore timely to study the potential of high energy neutrino 
and photon observations for scrutinizing the correlation signal suggested 
by the Auger collaboration.

The idea that neutrinos and photons are produced as secondaries in 
CR interactions close to the core of AGN has a long history~\cite{agn}.  
In particular the diffuse neutrino and photon fluxes from all AGN
have been studied in great detail~\cite{agn2,agn1,agn3}. Moreover, the 
expected neutrino and photon fluxes from Cen~A were discussed recently  
in view of the PAO results in Refs.~\cite{Cuoco:2007qd,Halzen:2008vz}
and \cite{Gupta:2008tm}, respectively. The present work extends these 
later studies by calculating both neutrino and photon secondary fluxes 
and by modeling the source and particle interactions in more detail. 
Acceleration of protons to energies as high as $10^{20}$\,eV in Cen~A is 
used as an assumption that we discuss only as far as the 
source parameters are concerned.

\section{Source and acceleration models for Cen A}

\subsection{Source parameters, the primary photon field and the
gas column density}

\paragraph{Accretion}
A general review of the properties of Cen~A, that is classified as FR~I 
radio galaxy and as Seyfert~2 in the optical, is given in 
Ref.~\cite{Israel:1998ws}. Its observed  spectral energy 
distribution (SED) of electromagnetic radiation is discussed in detail by
Ref.~\cite{steinle06}. Although Cen~A is  the nearest active 
galaxy with its distance\footnote{At this distance $1^{\prime\prime}=18$\,pc.}
of $D=3.8\,$Mpc~\cite{Rejkuba:2003xt}, 
there are a number of difficulties in
deducing the photon distribution  $n_\gamma(\vx,\vp)$ that serves as 
target for hadronic interactions from observational data. First, the 
resolution of e.g.\ X-ray 
satellites  like Chandra ($0.5^{\prime\prime}$) or XMM-Newton ($5^{\prime\prime}$)
is not sufficient to resolve the 
core of the AGN. Second, part of the emitted radiation is heavily absorbed 
by the dust lane hiding the AGN core. Third, it is useful to distinguish 
between primary photons that serve as targets for hadronic interactions and 
secondary photons produced therein. Finally, the observed photon flux
is at most energies probably dominated by photons produced 
in purely electromagnetic processes.

We prefer therefore to model the primary photon field around the AGN core 
guided by the simplest possible theoretical model~\cite{sunyaev}. The 
thermal emission from a geometrically thin, optically  thick Keplerian 
accretion disc is described by the temperature profile 
\be \label{T}
 T(r) = \left( \frac{3GM\dot M}{8\sigma\pi r^3}
                \left[ 1-(R_0/r)^{1/2} \right] \right)^{1/4} \,.
\ee
Within this simple model, the mass $M$ of the central supermassive 
black hole (SMBH), its 
accretion rate $\dot M$ and the inner edge of the accretion disc $R_0$
fix the main part of the primary electromagnetic radiation. 
Recent estimates for  $M$ vary between
$M= (0.5-2)\times 10^8M_\odot$~\cite{mass}. 
We will use $M=1\times 10^8M_\odot$ and thus the Schwarzschild 
radius is $R_s=3\times 10^{13}\,$cm. The angular momentum of the
SMBH in Cen~A is not known, so we use as smallest radius
of the acceleration and of the emission region the radius of the last stable 
orbit for a Schwarzschild black hole, $R_0= 3R_s\ap 1\times 10^{14}\,$cm. 

The accretion rate $\dot M$ and accretion efficiency $\eta=L/(\dot M c^2)$ were 
fitted in Ref.~\cite{Evans:2004ev} to Chandra and XMM-Newton observations 
assuming spherical (Bondi) accretion. The obtained accretion rate 
$\dot M=6\times 10^{-4}M_\odot/$yr should be considered as a lower limit, 
because it does not account for the accretion of gas too cool to emit X-rays.
In particular, it has been argued in 
Refs.~\cite{Hardcastle:2007iz,Markowitz:2007qj} that the heavy X-ray 
absorption and the high metallicity indicate accretion of cold gas.

Integrating the surface brightness $D=\sigma T^4$ over the finite accretion 
disc for  the chosen values of $M,\dot M$ and $R_0$ reproduces the 
characteristic blue bump~\cite{bluebump}. Because of the relatively 
small accretion rate, the bump is shifted somewhat to lower 
frequencies compared to the standard case. The dust lane of Cen~A prevents
that such a bump can be seen by us directly in the SED of Cen~A, but 
the observed HII line emission that is easiest explained by UV irradiation 
from the the central nucleus is indirect evidence for its 
existence~\cite{Mouri94}.

Since the surface brightness drops fast with the radius, $D\propto r^{-3}$, 
most of the radiation is emitted close to the core, (3--15)\,$R_s$, cf. 
also Ref.~\cite{netzer}. To simplify our simulation, we consider therefore 
the following one-dimensional model for the source: We describe the 
accretion disc as a sphere of radius $R_1=15R_s$ filled with an homogeneous, 
isotropic photon field radiating photons with the same spectrum 
$n_\gamma(\eps)$ from each point on a ``photosphere'' with radius 
$R_1$~\cite{BLR}. We model the energy-dependence of $n_\gamma$ by 
\be \label{bump}
 n_\gamma(\eps) = K_{\rm UV} \left\{
           \begin{array}{ll} 
            25\,\eps_{\rm eV}          & \qquad\eps_{\rm eV} < 0.2 \\
            \eps_{\rm eV}^{-1}    & \qquad 0.2<\eps_{\rm eV} <5 \\
             0.1\,\eps_{\rm eV}^2\exp(-\eps_{\rm eV}/2)    
             & \qquad\eps_{\rm eV} > 5 \\
           \end{array}
           \right.
\ee
with $\eps_{\rm eV}\equiv \eps/{\rm eV}$. The exponential cutoff is 
connected to the maximal temperature of the disc close to $R_0$. 
Finally, we assume that a hot corona produces an additional X-ray component,
\be
 n_X(\eps) = K_{\rm X} \eps_{\rm eV}^{-1.7} \,,
\ee
where the exponent $-1.7$ is chosen to agree with the X-ray observations from 
Refs.~\cite{Evans:2004ev,Markowitz:2007qj} and we use as high-energy cutoff 
$\eps=100\,$keV.

\paragraph{Normalization}

The accretion rate $\dot M=6\times 10^{-4}M_\odot/$yr determines together 
with the accretion efficiencies $\eta_i$ the luminosity $L_i$ in the 
wave-length range $i$,  $L_i= \eta_i\dot M c^2$. For the X-ray range, we 
use $L_X=4.8\times 10^{41}$erg/s in the 2-10\,keV range according to the 
observations~\cite{Evans:2004ev,Markowitz:2007qj} together with 
\be
 L_i = \pi R_1^2\: c\int\d\eps\:\eps n_\gamma(\eps)=\eta_iL_{\rm Bondi}
 =\eta_i\dot Mc^2 \,,
\ee
giving $K_X=8\times 10^{11}/({\rm cm}^3\,{\rm eV})$ and $\eta_X=1.5\%$. 
The $\gamma$-ray luminosity $L_\gamma=5\times 10^{42}$\,erg/s observed by 
COMPTEL and OSSE~\cite{CGRO} corresponds to $\eta_\gamma=15\%$.
Finally, we choose the normalization of the UV bump, 
$K_{\rm UV}$ in Eq.~(\ref{bump}) 
as $\eta_{\rm UV}=10\%$.

The used numerical values of the various normalization constants are
summarized in Table~\ref{norm}. There we report also the resulting 
efficiencies $\tilde\eta=L_i/L_{\rm Edd}$ relative to the Eddington luminosity,
$L_{\rm Edd}= 1.5\times 10^{46}\,$erg/s, of a $10^8M_\odot$ black hole.
The combined value of $\sum_i\tilde\eta_i$ is in the range of 
advection-dominated accretion 
and thus the formation of a geometrically thin, optically thick accretion 
disc for Cen~A is not guaranteed. We will nevertheless assume the existence of
such a standard Shakura-Sunyaev accretion disk, keeping in mind however 
that the derived values for the photon density should be considered
as upper limits.

\begin{table}[bh]
\begin{center}
\caption{\label{norm}
The luminosities, efficiencies and normalization constants in different 
wave-length ranges.}
\medskip
\begin{tabular}{|c||c|c|c|c|}
\hline 
band & $L_i$\,/erg/s & $\eta$ & $\tilde\eta$ & $K_i/($cm$^3$\,eV)
\\  \hline
$\gamma$-ray &$5.0\times 10^{42}$& 15\% & $3.3\times 10^{-4}$ & --
\\ \hline
X-ray & $4.8\times 10^{41}$ & 1.5\% & $3.3\times 10^{-5}$ & $8\times 10^{11}$ 
\\ \hline
UV & $3.6\times 10^{42}$ & 10\% & $2.4\times 10^{-4}$ & $1\times 10^{13}$
\\ \hline
\end{tabular}
\end{center}
\end{table}


For large distances, $r\gg R_1$, the photon field emitted from the photosphere 
of radius $R_1$ scales as $n_\gamma(r)\propto (R_1/r)^2$ and has approximately
a distribution of momentum vectors with 
$\cos\theta={\bf p}\cdot {\bf r}/(|{\bf p}||{\bf r}|)\geq 1 -(R_1/r)^2$.
Thus we have to distinguish two cases: Either photons and/or cosmic rays 
are distributed isotropically or both are non-isotropic. The latter case
is realized when charged particles propagate along the (regular) 
field lines and primary photons stream nearly radial outwards for 
$r\gg R_1$. 
Above threshold $E\sim 10^{16}$\,eV, the resulting interaction depth is 
$\tau_{p\gamma}\sim 3$.

The PAO data used in the analysis~\cite{pao_corr} correspond to an
exposure $\Xi=9000\,$km$^2$\,yr\,sr with maximal zenith angle
$\theta_{\max}=60^\circ$. The correlation signal with AGN was
maximized for the threshold energy $E_{\rm th}=6\times 10^{19}\,$eV
and the angular bin size $3.1^\circ$. For these values, two events
were found in the angular bin around Cen~A. Centaurus~A is close
enough to the Earth that energy losses of CRs with energy close to
$E_{\rm th}$ can be neglected, cf.\ e.g.\ Fig.~1 of
Ref.~\cite{Kachelriess:2007bp}.
Assuming that both events indeed originate from 
this AGN, the integral CR flux above $E_{\rm th}$ on Earth from Cen A is
\be  \label{F}
 F(>E_{\rm th}) = \frac{2\,\Omega}{R\,\Xi}
 = 2 \times 10^{-3} \,\frac{1}{\rm{km^2 \, yr}} \,.
\ee
Here, $\Omega\ap 9$\,sr is the field-of-view of the PAO and 
$R=\eta(\delta_s)/\langle\eta(\delta)\rangle\ap 0.95$ is the ratio of the 
exposure $\eta(\delta_s)$ at the declination $\delta_s=-43.0^\circ$ 
of Cen~A and the average exposure $\langle\eta\rangle$.

\paragraph{Gas column density  and proton-proton interactions}

Reference~\cite{Markowitz:2007qj} fitted Suzaku observations of Cen~A with 
several absorption models. The primary X-ray component was found to be 
absorbed by the column density $X=1\times 10^{23}/$cm$^2$, while a less 
brighter component is more heavily absorbed with $X=7\times 10^{23}/$cm$^2$. 
These results can be interpreted either as indication for a clumpy structure 
of the gas or for two different $X$-ray sources, e.g.\ from accretion and 
from jet emission. The resolution of these observations is however rather 
limited, and the value for the column density is therefore biased by the 
dense torus.
A more representative determination of the mean density of the gas around 
the core of Cen~A was possible with XMM-Newton and Chandra observations.
Ref.~\cite{Kraft:2003gp} determined the density of the interstellar
medium around the core of Cen~A as $n_H=n_0[1+(r/r_0)^2]^{-0.6}$ with
$n_0=0.04/$cm$^3$ and $r_0=0.5$\,kpc.
Finally, Ref.~\cite{Worrall:2007tf} found as average hydrogen density
$X=1.5\times 10^{21}/$cm$^2$ along the radio jet starting from a projected 
distance of 0.3\,kpc up to 2.5\,kpc. With $d=0.4$\,kpc as diameter
of the jet and assuming that the jet axis is close to perpendicular to the
line-of-sight~\cite{jetaxis}, an average density $n_H\ap X_H/d \ap 1.7/$cm$^3$
follows.

\subsection{CR acceleration, propagation and interactions in the source}

X-ray observations of Cen~A allow one to trace the acceleration sites
of electrons along  the radio jet, because their synchrotron loss length 
is short compared to the extension of the jet. The strong dependence
of synchrotron emission on the mass of the radiating particle implies that 
the SED is, apart from the highest energies, dominated by electromagnetic 
interactions of electrons and that therefore acceleration site and mechanism 
for electrons and protons may differ. Although various proposal for the 
acceleration of protons to UHE exist, we restrict ourselves to two basic 
models, namely shock acceleration in the radio jet and acceleration in the  
regular electromagnetic field close to the core of the AGN. These two models 
are characterized by a rather different set of parameters and the secondary 
fluxes in several other models may be  obtained by ``appropriate 
interpolation.'' 
We neglect relativistic effects, because of the moderate Lorentz factors
observed in Cen~A and the large angle between the jet axis and the 
line-of-sight.

\paragraph{Acceleration in regular fields near the core}

Acceleration close to the core could proceed either via shock acceleration 
in accretion shocks~\cite{KE86} or via acceleration in regular fields. 
The former case is disfavoured since the acceleration rate is smaller 
than the rate of photonpion energy losses~\cite{PhTr}. Therefore we
consider here only acceleration in regular fields close to the core. 
Among the possibilities discussed in the literature are electromagnetic 
winds from the BH/disk magnetosphere \cite{Bl76,Lo76},
and unipolar induction around a Kerr BH \cite{BZ77,MT82}.
The regular magnetic field close to the core consists of a toroidal component 
in the accretion disc and a poloidal field component $B_p=B_0(R_0/r)^\beta$.
For field strengths $B_0\sim $\,kG acceleration to energies around 
$(10^{20}$\,eV is possible. Since the curvature radius $R_C$ of the 
field-lines is typically large, curvature radiation of protons is not
very effective. Moreover, protons move mainly parallel to the field-lines 
and thus also synchrotron losses are suppressed. Because of our simplified
one-dimensional geometry, we assume that the combined effect of synchrotron
and curvature radiation is such that is does not affect protons but acts 
as main energy loss for electrons. The other remaining free parameter in this
model, the injection point of the CRs, is fixed to $r=6R_s$.

The required magnetic field strength for acceleration to $10^{20}$\,eV
is an order of magnitude higher than one would expect from equipartition, 
$B^2/8\pi\sim L/4\pi R_0^2c$. The currently low accretion rate and small
luminosity of Cen~A disfavor therefore acceleration close to the core 
as  acceleration mechanism, if Cen~A was not in the past in a state of 
higher activity.

\paragraph{Acceleration in the radio jet}

Rachen and Biermann suggested the hot spots in FR-II radio galaxies as 
sites of UHECR acceleration~\cite{RB}. These spots are formed as
termination shock of the supersonic jets in the intergalactic medium 
and are especially prominent in FR II jets. They are instead dim or absent 
in weak FR I sources, most likely because strong turbulent dissipation 
in the propagation phase reduces the momentum finally released at the 
termination shock. Nevertheless, e.g.\ Ref.~\cite{Romero:1995tn} argued
that at the ``hot spot'' in Cen A acceleration of protons to UHE occurs.
The projected size of the hot spot is $R_{HS}\ap 1.7\,$kpc, containing
magnetic fields with strengths $B\ap 0.5\:$mG. 
Alternatively, proton could be accelerated along the whole extension of 
the radio jet, extending  a projected distance of 6\,kpc from the 
nucleus. In this case, protons have to diffuse
through the X-ray photons emitted by accelerated electrons and, more
importantly, through the hydrogen gas observed in the jet. 

Thus we shall use acceleration in the jet as our second basic scenario. 
More precisely, we consider as acceleration 
region for protons a cylinder of length $l=4\,$kpc and diameter $d=0.3\,$kpc,
similar to the emission volume of X-rays observed by Chandra. Diffusion
in the turbulent magnetic fields  will increase the interaction depth.
We choose as field strength $B=0.2$\,mG, set the coherence length
conservatively equal to the length $l_c=1$\,kpc and use $a=1/2$ for the
energy dependence of the diffusion coefficient, $D(E)\propto E^{a}$,
corresponding to a Kraichnan fluctuation spectrum.
The choice $a=1/2$ is intermediate between the often assumed Bohm and
Kolmogoroff diffusion as well as close to the numerical results of
Ref.~\cite{lemoine}.

\paragraph{Relative importance of photon and proton targets}

In the case of acceleration close to the core, either in accretion shocks
or regular electromagnetic fields, UV photons are the most important
scattering targets and the interaction depth for photo-hadron interactions
can reach $\tau_{p\gamma}\sim $~few.
In all other cases, $\tau_{p\gamma}$ will be reduced at least by a factor 
$\sim 10R_s/l$ where $l$ is the typical dimension of the acceleration region.
Diffusion increases the effective size of the source, but this effect becomes
noticeable mainly at energies below the threshold for photon-hadron 
interactions. As a rule of thumb, photo-hadron interactions are therefore 
only important when the acceleration takes place close to the core.

The importance of proton-proton interactions depends more strongly on the 
concrete model assumptions -- as the scatter in the observed column density 
$X$ of gas shows. Moreover, the accretion rate may be time-dependent and 
matter close to the core is concentrated in clumps. Thus both the CR 
luminosity $L_{\rm CR}$ and the interaction depth $\tau_{pp}$ can vary 
significantly depending on the model assumptions. While the importance
of gas as target close to the core is uncertain, the Chandra observations
indicate that pp scattering is the main source of CR interactions in the
jet.

\paragraph{Energy spectrum and interactions}

We consider three cases for the generation spectrum $dN/dE\propto E^{-\alpha}$
of UHECRs: $i)$ A power-law with $\alpha\ap 2$ as conventionally predicted
from  first-order Fermi acceleration with non-relativistic shocks.
$ii)$ A broken power-law with break energies $E_b=10^{17}\,$eV and 
$E_b=10^{18}\,$eV, respectively, and $\alpha = 2.7$ for $E>E_b$ as suggested 
by the dip-interpretation of
the experimental data~\cite{dip}. Since the steepening of the observed diffuse
spectrum may result from a distribution of maximal energies 
${\rm d} n/{\rm d}E_{\max}$~\cite{KS06}, present UHECR observations do not 
distinguish between these two possibilities, even if one assumes that 
extragalactic sources dominated the flux down to $10^{17}$\,eV. 
Note also that $\alpha=2.7$ is consistent with the observed radio spectrum
of most AGN. In order to weaken the energy demands of Cen~A,
we extend the $\alpha= 2.7$ spectrum with $\alpha=2$ below $E_b$.
$iii)$ A flat spectrum with $\alpha=1.2$ and most energy concentrated at 
$E_{\max}$, as expected for a ``linear accelerator.''
In all three cases we use $E_{\max}=10^{20}\,$eV.

Hadronic interactions are simulated with an extension of the Monte Carlo code
described in Ref.~\cite{II}, including the possibility of a non-thermal and 
anisotropic photon field $n_\gamma({\bf x},{\bf p})$ and of proton-proton
interactions. Electromagnetic processes\footnote{A description 
of our treatment of electromagnetic cascades and results will be presented 
elsewhere.} like pair production and inverse 
Compton scattering of electrons and photons both inside the source and on 
IR, cosmic microwave background (CMB) and radio photons are also taken into 
account. 
Finally, diffusion is simulated as described in Ref.~\cite{II}.

\section{Cosmic ray, photon and neutrino fluxes}

\begin{figure}
\begin{center}
\epsfig{file=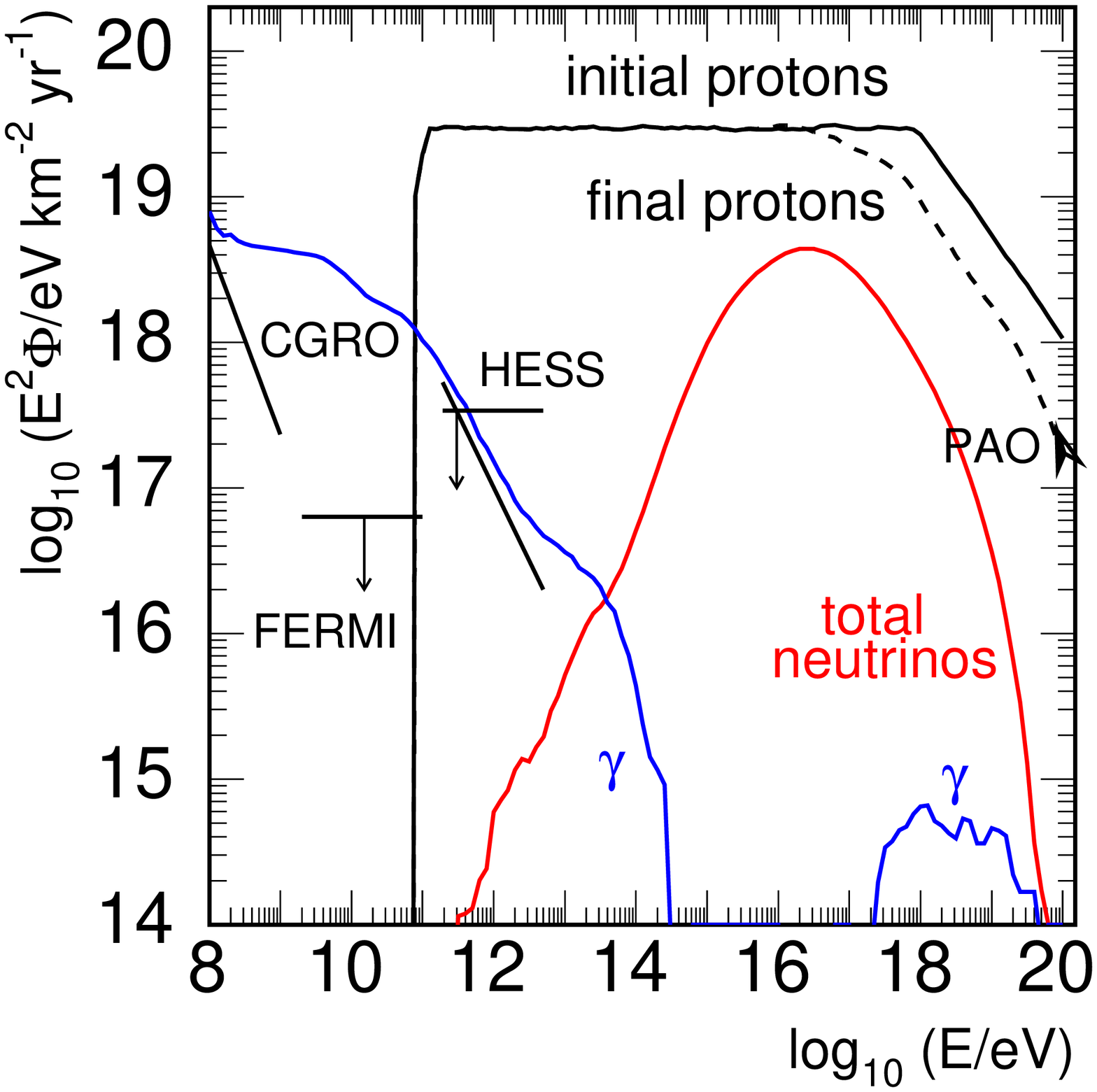,width=0.43\textwidth,angle=0}
\epsfig{file=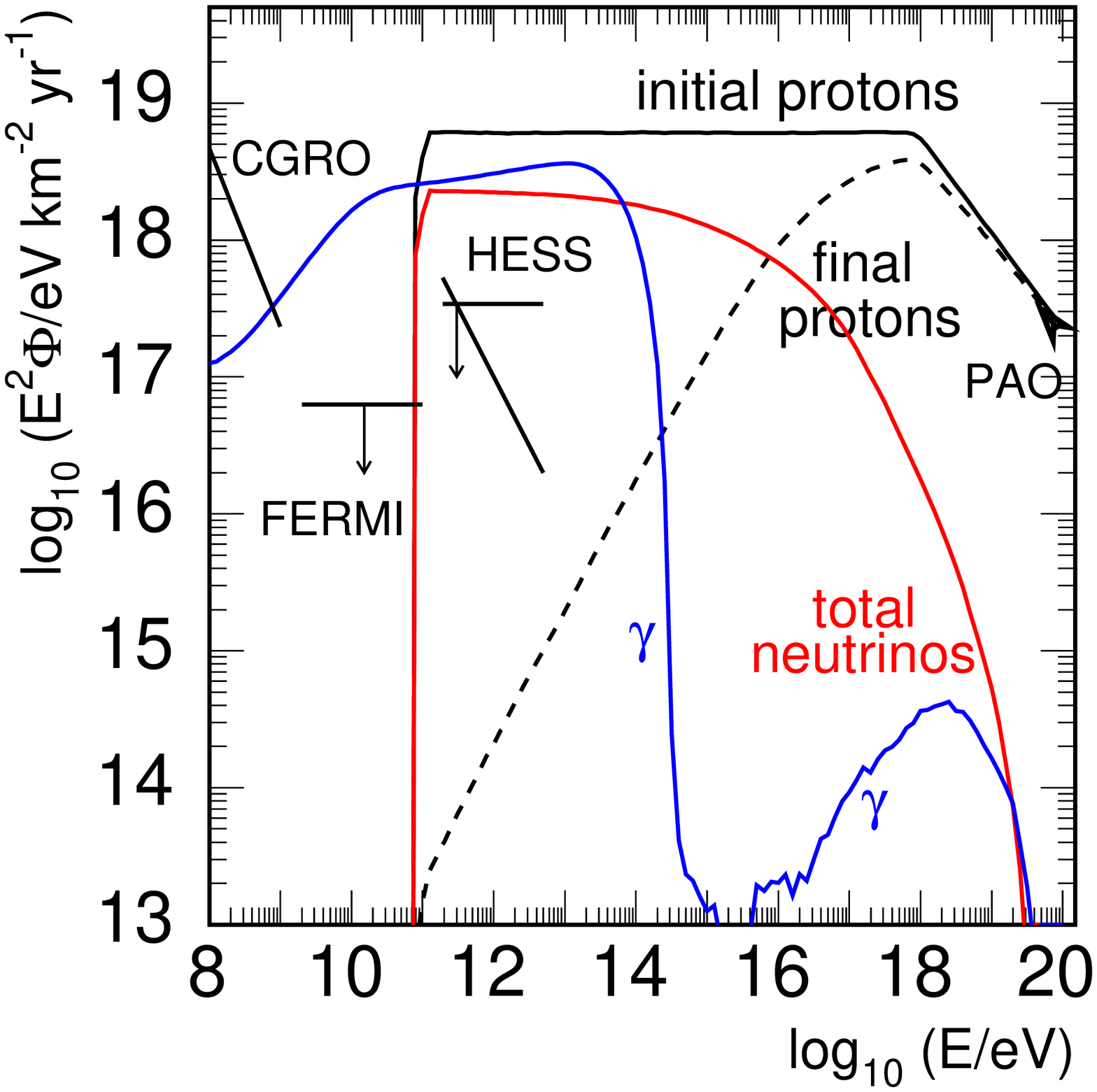,width=0.43\textwidth,angle=0}
\end{center}
\vspace*{-.7cm}
\begin{center}
\epsfig{file=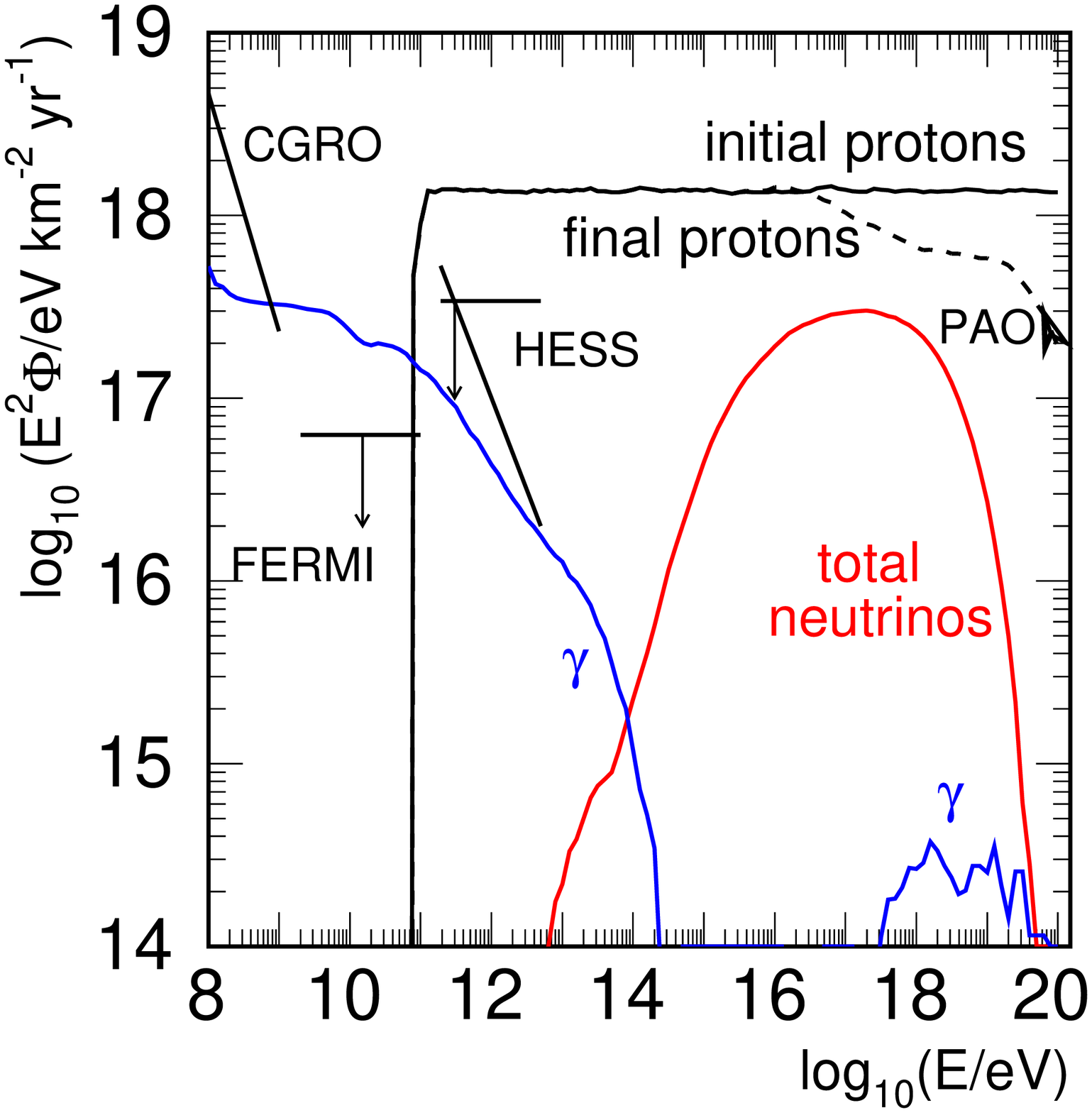,width=0.43\textwidth,angle=0}
\epsfig{file=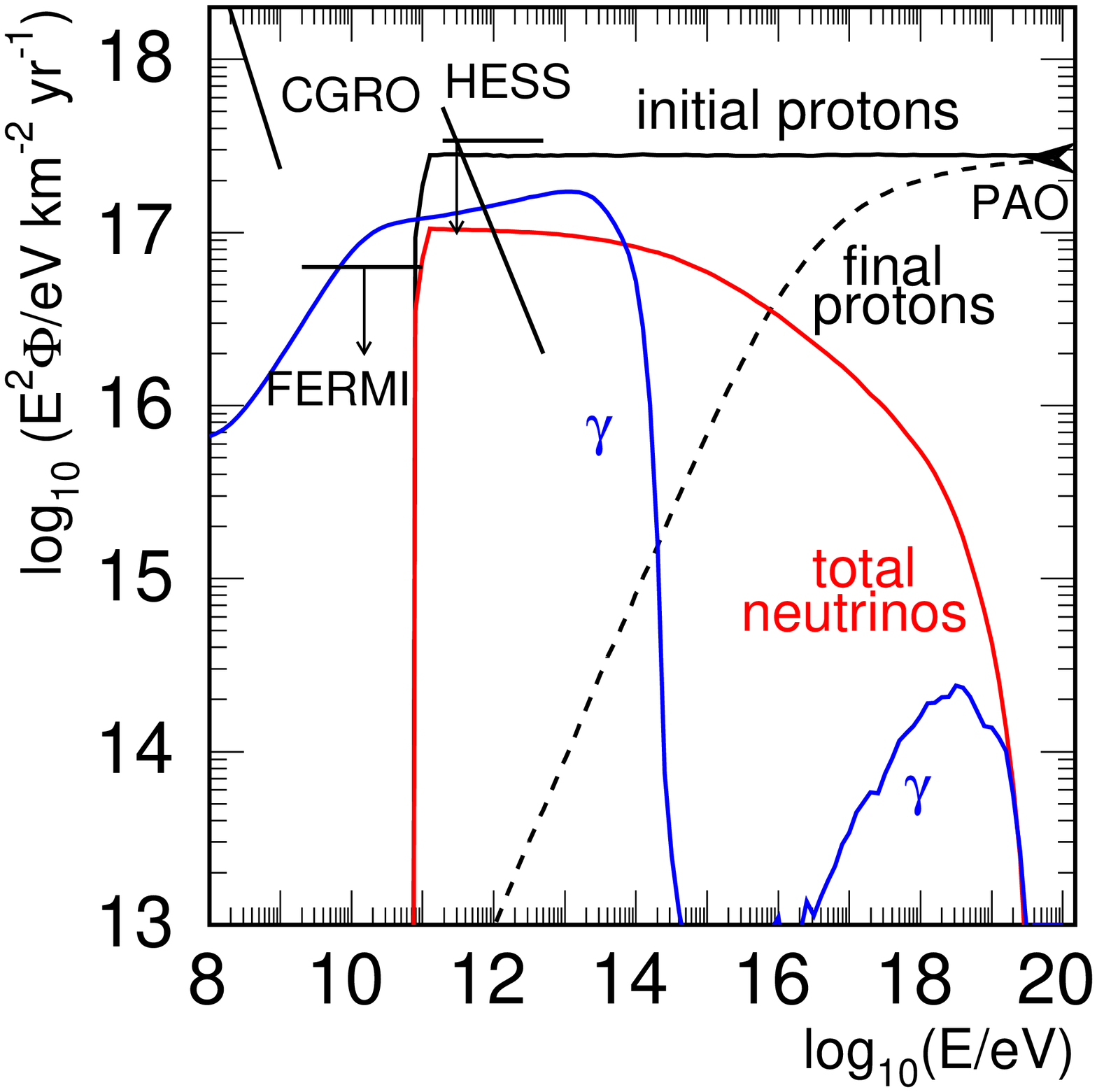,width=0.43\textwidth,angle=0}
\end{center}
\vspace*{-.7cm}
\begin{center}
\epsfig{file=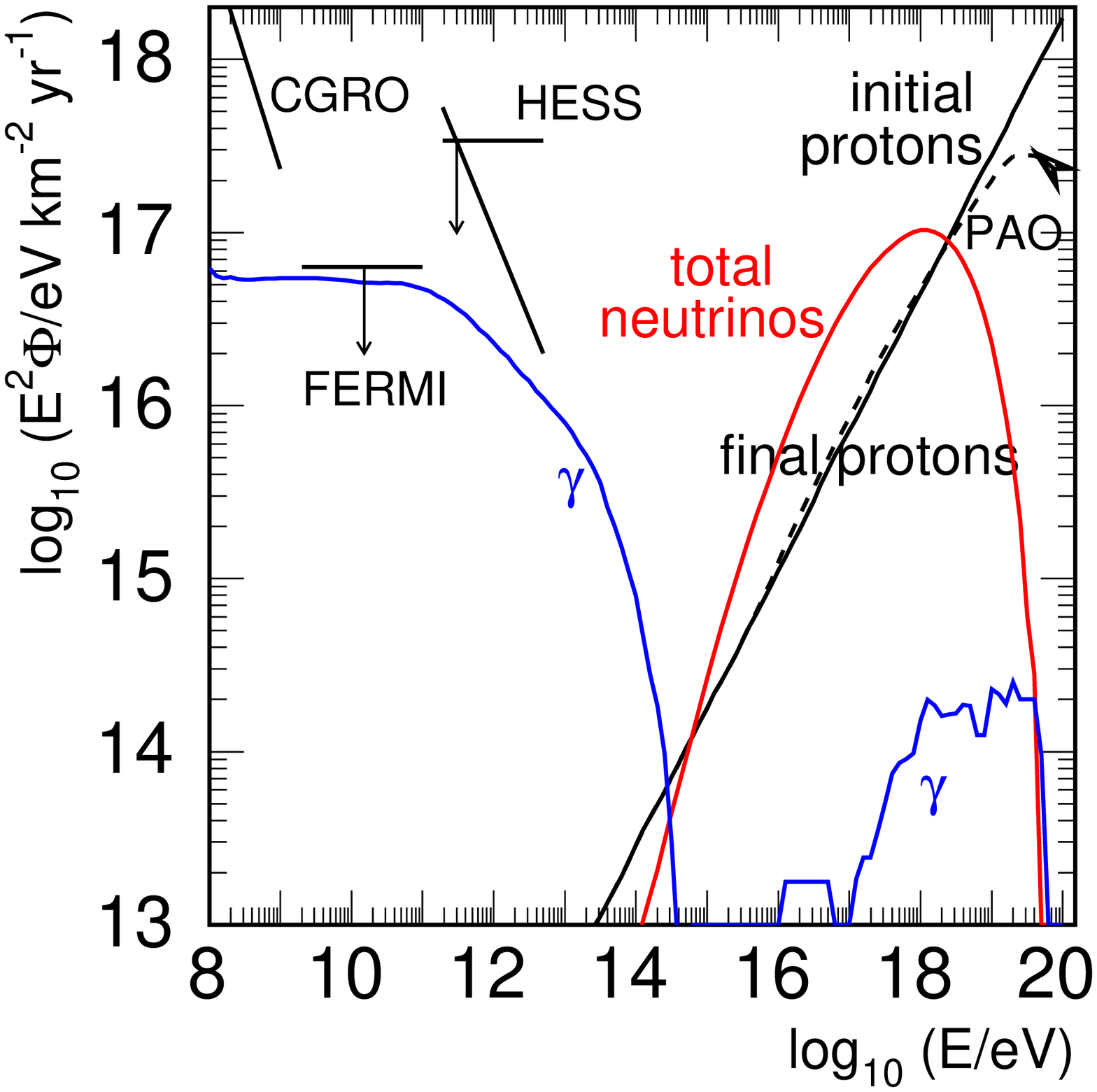,width=0.43\textwidth,angle=0}
\epsfig{file=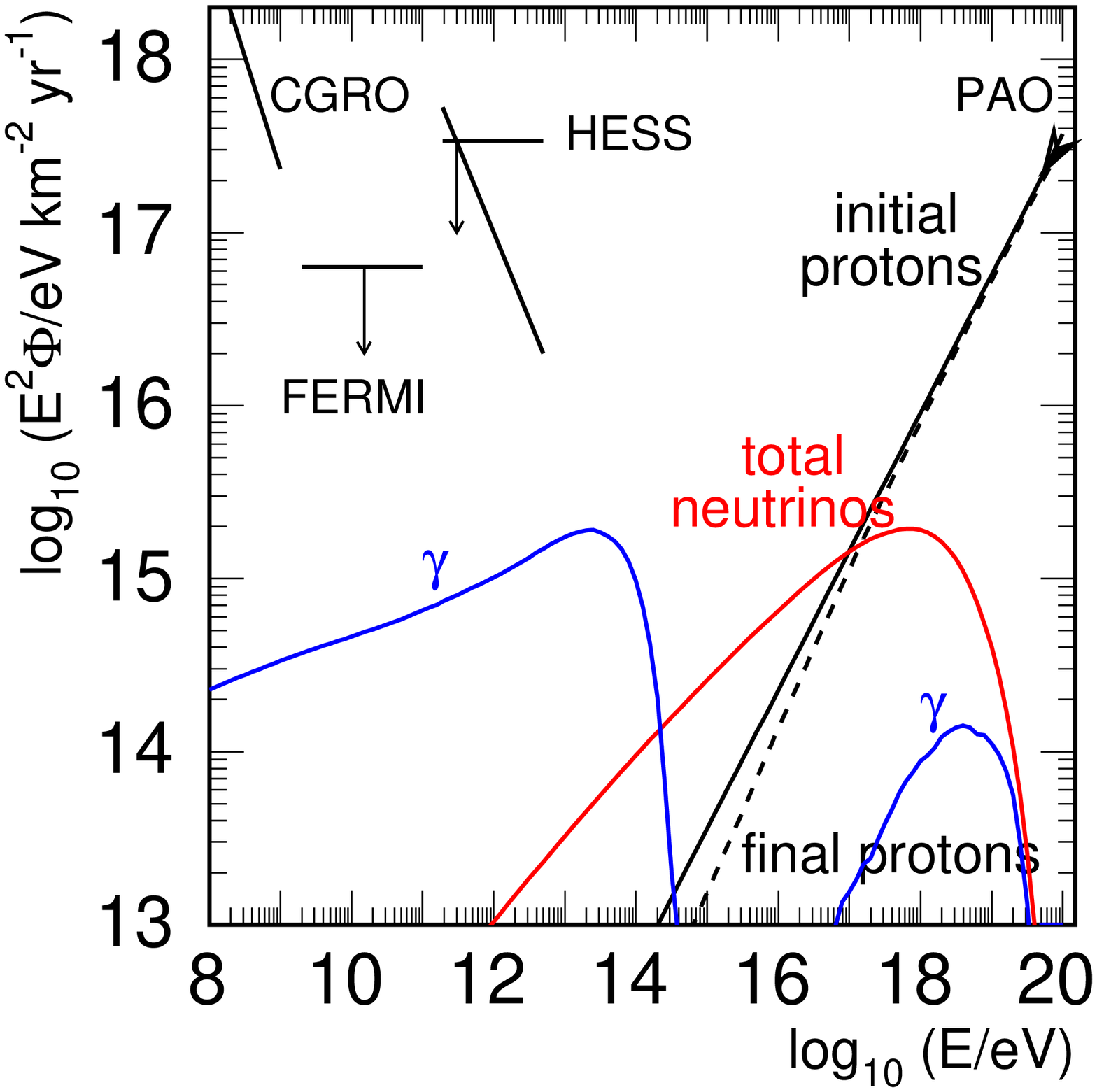,width=0.43\textwidth,angle=0}
\end{center}
\caption{Particle fluxes $F(E)$ from Cen~A  normalized to the PAO data as 
function of the energy $E$: 
top panels broken power-law with $\alpha = 2$ and 2.7 and break energy 
$E_b=10^{18}$\,eV, 
middle $\alpha=2$, and bottom panels $\alpha=1.2$;
left panels acceleration close to the core, right panels in the
jet;
initial protons (solid black), final protons (dashed black), 
photons (blue), and sum of all neutrinos (red). 
\label{spec1}}
\end{figure}

The left and right panels of Fig.~\ref{spec1} display the particle 
fluxes $F(E)$ predicted from Cen~A  as function of the energy $E$ in the case 
of acceleration close to the core and in the jet, respectively. 
Additionally to the injected proton flux (black solid line), we show 
the flux of protons (black dashed)  arriving on 
Earth. %
The final proton flux is reduced by interactions by a factor 
$\ap 2$ above the threshold energy $\sim 10^{16}$\,eV
for photon-proton interactions (left), while diffusion in the jet increases
the interaction depth for lower energies (right panels), 
resulting in the effective production of secondaries.

The photon flux at Earth after cascading in the source and on the 
extragalactic background light (EBL) consists of two contributions: $i)$ 
Photons produced in the source, which survive their subsequent travel 
to the observer without interactions with the EBL; $ii)$ photons produced 
during the cascade on the EBL. While the former are always observed as a 
pointlike contribution from a pointlike source, the latter are characterized by 
a finite angular spread, due to the deflections of cascade electrons in 
the extragalactic magnetic field (EGMF).
Depending on the strength of the latter and on the experimental resolution, 
the second contribution may be observed as a halo 
around the source~\cite{halo}, while contributing only partially to 
the unresolved  pointlike flux. Because of the proximity of Cen~A,
deflections in the EGMF have a---compared to the overall 
uncertainties---negligible influence on the calculated
photon spectra. 

The photon flux at Earth after cascading in the source and on the ELB
is shown as a solid blue line together with the combined fit to 
EGRET, OSSE and COMPTEL observations of Cen~A from Ref.~\cite{CGRO}, 
the H.E.S.S. limit from Ref.~\cite{Aharonian:2005ar} for $\alpha=2$ and 
$\alpha=3$, and the FERMI sensitivity~\cite{glast} for a $5\sigma$ detection 
of a point source in 1~yr. The predicted spectrum shows the typical 
suppression above the pair production threshold on the CMB at 
$E\ap 200$\,TeV. Since the CR spectra are normalized to the integral UHECR 
flux above $E_{\rm th}= 5.6\times 10^{19}\,$eV, steeper spectra result in
larger secondary fluxes at low energies. In particular, the
photon flux overshoots the H.E.S.S. limit or the CGRO observations
in the case of a broken power-law injection spectrum. Thus already 
present observational data favour either softer UHECR 
spectra than $\d N/\d E\propto E^{-2.7}$ for Cen~A or a lower UHECR flux than
expected from the normalization to the Auger observations.
Note however that our normalization relies on only two events and has 
therefore a large statistical uncertainty. On the other hand, the 
determination of the energy scale of UHECR experiments is notoriously
difficult and it has been argued that the PAO energy scale
should be shifted up to obtain agreement with the spectral 
shape predicted by $e^+e^-$ pair production~\cite{dip}.

Comparing the neutrino flux at $10^{16}$\,eV for the different models allows 
one to judge how strong the predicted neutrino event number depends on 
the slope of the CR spectrum. Going from a broken power-law with 
$\alpha=2.7$ at UHE to $\alpha=1.2$ reduces between three and four orders 
of magnitude 
the neutrino flux at 100\,TeV. Calculating the expected event number 
in a neutrino telescope requires a definite choice of the experiment. 
Centaurus~A is from the location of Icecube only visible from above, and 
thus the background of atmospheric muons allows only the use of contained 
events that carry essentially no directional information 
($\delta\theta\sim 30^\circ$). For the calculation of the number of contained 
events expected in Icecube we use as effective volume $V=1\,$km$^3$, as 
threshold $E_{\min}=100$\,TeV, assume 100\% efficiency above $E_{\min}$ and 
use the CTEQ5 neutrino-nucleon cross sections~\cite{cteq}. By contrast,
a neutrino telescope in the Mediterranean could make use of the muon signal
and the directional information. In this case the rate $R$ of muon events 
can be approximated by 
\be
 R= A \int_{E_{\min}}^\infty\d E\: S(E)F_{\nu_\mu}(E) P(E,E_{\min})
\ee
with $A=1\,$km$^2$, the probability 
$P(E,E_{\min})=N_A\langle R(E,E_{\min})\rangle\sigma_{\nu N}$ that a muon reaches 
with $E>E_{\min}$ the detector and the angular averaged shadowing factor 
$S(E_\nu)$ accounting for attenuation of the neutrino flux in the 
Earth~\cite{gandhi}. 
The resulting event numbers both for cascade and shower event number per 
year observation time are summarized in Table~\ref{table} together with the 
input CR luminosity. 
Note that the cutoff in the neutrino spectra below 100\,GeV that is visible
in the upper, right panels of Fig.~1 is artificial, since we neglect 
neutrinos with lower energies in our simulation. 

In summary, we predict that for all cases where there 
is a marginal chance to detect a neutrino signal from Cen~A, atmospheric
Cherenkov telescope and/or $\gamma$-ray satellites like FERMI should detect 
previously a photon signal from Cen~A.  Main reason for this result---that 
is in contradiction to earlier expectations, cf.\  e.g.\ 
Ref.~\cite{Neronov:2002xv}---is the cascading of photons in the anisotropic
photon field close to the source.

Finally we remark that we assumed for the calculation of all fluxes an 
isotropic emission. This is well justified in the case of stochastic 
acceleration in the jet, because of the small gamma factors of the 
observed matter flows in the jet and the large angle between the jet 
and the line-of-sight. In the case of acceleration close to the core,
protons move along the field lines and thus the emission is rather 
anisotropic. Since we used the UHECR flux towards our line-of-sight as 
normalization, the total UHECR luminosity of Cen~A is therefore larger than
the one estimated in Table~1. On the other hand, the predicted number of 
neutrino events is not affected by the anisotropy, since the
ratio of neutrino and UHECR fluxes is roughly independent from the
considered direction. By contrast, TeV photons as final products
of electromagnetic cascades are more effectively isotropized than
UHECRs. This effect will slightly increase the TeV photon flux compared
to the isotropic case assumed in Fig.~1.

\begin{table}[bh]
\begin{center}
\caption{\label{table}
The injected CR luminosity $L_{\rm CR}$ and the number of neutrino events
expected per year observation time for different energy slopes and acceleration 
scenarios.}
\medskip
\begin{tabular}{|c||c|c|c|c||c|c|c|c|} 
\hline
 &\multicolumn{4}{|c||}{jet}&\multicolumn{4}{|c|}{core}\\
\cline{2-9}
$\alpha$ or $E_b$/eV & 1.2 & 2.0 & $10^{18}$ & $10^{17}$ 
         & 1.2 & 2.0 & $10^{18}$ & $10^{17}$ 
\\ \hline
$L_{\rm CR}/10^{40}$erg/s & 0.35 & 0.97 & 5.2 & 5.2  
                        & 1.1 & 2.2 & 11 & 10
\\  \hline
contained $\# \;\nu$/yr & $8\times 10^{-5}$ & $0.02$ & 0.4 & 2.0
                        & $7\times 10^{-4}$ & $0.01$ & 0.3 & 0.9
\\  \hline
$\# \;\mu$/yr & $4\times 10^{-5}$ & $7\times 10^{-3}$ & 0.2 & 0.7
              & $3\times 10^{-4}$ & $7\times 10^{-3}$ & 0.1 & 0.5
\\ \hline
\end{tabular}
\end{center}
\end{table}

The expected neutrino and photon fluxes from Cen~A were discussed recently  
in view of the PAO results in a rather qualitative way
in Refs.~\cite{Cuoco:2007qd,Halzen:2008vz}
and \cite{Gupta:2008tm}, respectively. Our case of acceleration close to the 
core and a broken power-law corresponds roughly to the set-up of 
Ref.~\cite{Cuoco:2007qd}, but the results differ by the large factor 30. 
One possible explanation of this difference is that their neutrino flux drops 
faster than we found. Reference~\cite{Halzen:2008vz} assumed pp interactions
on gas close to the core with $\tau_{pp}\sim $~few as production mechanism.
The obtained event numbers agree well taking into account the differences
in the used assumptions.
Finally, Ref.~\cite{Gupta:2008tm} discussed photons from hadronic 
interactions in Cen~A. This work did not include the effect of 
electromagnetic cascading and the conclusions are therefore difficult to 
compare.

\section{Conclusions}

We have calculated the flux of high energy cosmic rays and of accompanying 
secondary photons and neutrinos expected from Cen~A. We  modeled
the distribution of target photons and gas guided by the simplest 
theoretical model for the accretion disc and by observational data, 
respectively.  The production of secondaries and the electromagnetic 
cascading of electrons and photons was simulated with a Monte Carlo 
procedure. 

In contrast to previous works, we showed
that scattering on gas becomes important if acceleration takes place along 
the jet. Moreover we found that a source that has an interaction depth
$\tau_{p\gamma}\gsim 1$ can be observed in the (1--100)\,TeV range by
atmospheric Cherenkov telescopes. Additionally to these more technical
results, we have shown that a combination of the old CRGO observations and
the limits from atmospheric Cherenkov telescopes can be used to constrain 
currently favoured UHECR models. In particular, we found that these data
favour either a softer UHECR injection spectrum than 
$\d N/\d E\propto E^{-2.7}$ for Cen~A or a lower UHECR flux than
expected from the normalization to the Auger observations.

One should remind however the main underlying uncertainties interpreting our 
results: The normalization of all fluxes is based on the assumption that 
two UHECR protons in the PAO data originate  from Cen~A. 
Note that several authors have argued that a larger number of events
originates in Cen~A~\cite{many}.
Apart from the purely 
Poisson error, the normalization may be influenced  by deflections in (extra-) 
galactic magnetic fields, the uncertainty in the energy scale of PAO, and a 
possible admixture of heavy nuclei. Deflections of UHECRs result in their
delayed arrival with respect to photons and neutrinos, introducing an additional
source of uncertainty in their relative normalization. Moreover, our model 
parameters ($M$, $\dot M$, $\eta_{\rm UV}$,\ldots) and even as 
basic parameters as the distance to Cen~A have sizeable uncertainties. 
Last but not least we have made several
simplifying assumptions like the use of an one-dimensional geometry and the 
omission of the acceleration process: We only {\em postulated\/} that
acceleration to $10^{20}$\,eV  is possible in the environment of Cen~A,
without demonstrating it for a concrete model. Despite these drawbacks, it is 
remarkable that
$\gamma$-ray and TeV observations of Cen~A allow one already now to constrain 
currently favoured UHECR models. The potential of neutrino telescopes to 
observe Cen~A depends strongly on the steepness of the UHECR generation 
spectrum. A neutrino telescope on the northern hemisphere would be very
useful for this task.

\section*{Acknowledgments}
We would like to thank the anonymous referees, Dan Evans and Elisa Resconi 
for their useful comments and Alessandro Cuoco for discussions that initiated 
this work.
S.O.\ and R.T.\ acknowledge support from the Deutsche Forschungsgemeinschaft 
within the Emmy Noether program and SFB 676, respectively.


\end{document}